%% file: paper.tex
\documentclass{vgtc}                          

\graphicspath{{figures/}{pictures/}{images/}{./}} 

\usepackage{times}                     

\usepackage{tabu}                      
\usepackage{booktabs}                  
\usepackage{lipsum}                    
\usepackage{mwe}                       

\usepackage{mathptmx}                  

\newcommand{\rev}[1]{#1}

\onlineid{0}

\vgtccategory{Research}

\vgtcinsertpkg

\title{Investigating Encoding and Perspective \\ for Augmented Reality Motion Guidance}

\author{Jade Kandel\thanks{e-mail: \{kandelj, spiros, dszafir, fuchs, dnszafir\}@cs.unc.edu}
\and Sriya Kasumarthi\thanks{e-mail: sriyakasumarthi@gmail.com} \\ %
\and Spiros Tsalikis $^*$\\ 
\and Chelsea Duppen\thanks{e-mail: \{chelsea\_parker, mlewek\}@med.unc.edu}\\ %
\and Daniel Szafir  $^*$~~~~~\ 
Michael Lewek $\ddag$ ~~~~~\
Henry Fuchs $^*$~~~~~\
Danielle Szafir $^*$ \\ %
     \scriptsize University of North Carolina at Chapel Hill\\
     }

\teaser{
  \centering
  \includegraphics[width=\linewidth]{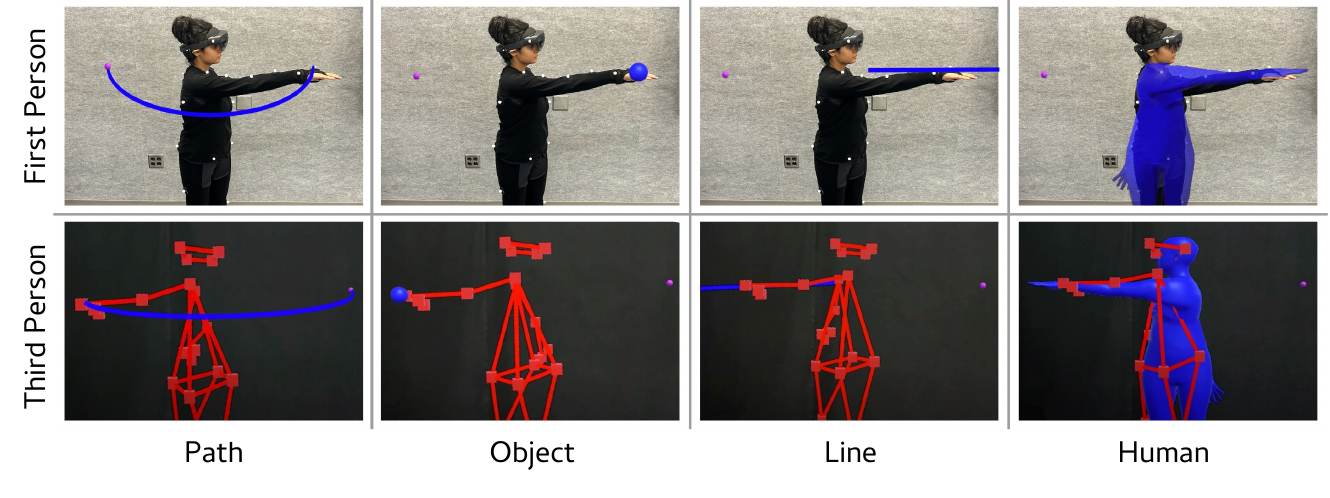}
  \caption{We 
  measured the effects of visualization techniques on accuracy and usability for guiding motion. Participants used 
  8 visualizations (4 encodings $\times$ 2 perspectives\rev{, seen above with trunk rotation}) to complete 3 isolation exercises. 
  The encodings 
  cued motion using \textit{path},  \textit{object},  \textit{line}, and  \textit{human}. 
  Perspectives included either the first- (top) or third-person (bottom) perspective.  }
  \label{fig:teaser}
}

\abstract{

Augmented reality (AR) offers promising opportunities to support movement-based activities, such as personal training or physical therapy, with real-time, spatially-situated visual cues. While many approaches leverage AR to guide motion, existing design guidelines focus on simple, upper-body movements within the user's field of view. We lack evidence-based design recommendations for guiding more diverse scenarios involving movements with varying levels of visibility and direction. We conducted an experiment to investigate how different visual encodings and perspectives affect motion guidance performance and usability, using three exercises that varied in visibility and planes of motion. Our findings reveal significant differences in preference and performance across designs. Notably, the best perspective varied depending on motion visibility and showing more information about the overall motion did not necessarily improve motion execution. We provide empirically-grounded guidelines for designing immersive, interactive visualizations for motion guidance to support more effective AR systems.

} 


\begin{document}
\maketitle


\input{1_Introduction}
\input{2_Related_Work}
\input{3_Motivation_and_Hypotheses}

\input{4_Methods}

\input{5_Results}
\input{6_Discussion}

\input{7_Limitations_and_Future_Work}

\input{8_Conclusion}


\acknowledgments{This work was supported by the National Institutes of Health Award 1R01HD111074-01 and NSF-IIS \#2320920.}
\bibliographystyle{abbrv-doi}

\bibliography{template}
\end{document}

%% file: 1_Introduction.tex
\section{Introduction}


Augmented Reality (AR) can guide motion by situating feedback with respect to a person’s body, improving spatial understanding and performance 
with direct visual guidance \cite{bressa_whats_2022}. Compared to traditional video and photos, AR also enhances proper body positioning in 3D space through better spatial perception of depth, height, and size by integrating binocular cues \cite{marriott_immersive_2018, whitlock_graphical_2020, kraus_value_2021}. Previous research 
uses AR for 
guiding motion in applications such as mechanical tasks \cite{gemito_mixed_2023},
medical training \cite{heinrich_comparison_2019}, 
sports \cite{lin_towards_2021, kaplan_towards_2018, han_my_2017, hoang_onebody_2016, jo_flowar_2023,anderson_youmove_2013}, and rehabilitation \cite{tang_physiohome_2015, ayoade_novel_2014, yu_perspective_2020}. However, few empirically-grounded recommendations exist for effectively designing immersive visualizations for motion guidance. 
Studies for guiding motion with AR \rev{have largely} focused on frontal plane upper body motion \cite{hoang_onebody_2016, yu_perspective_2020}; however, these findings may not generalize to other planes or types of motion. \rev{Further}, while previous studies have compared various visualization methods for representing motion trajectories \cite{tang_physiohome_2015,durr_eguide_2020, yu_perspective_2020}, we lack insight into how these paths compares to alternative methods.
For example, virtual human guides or discrete targets more closely mirror 
real-world motion guidance, such as mimicking an instructor's demonstration or reaching for a physical target. 
Comparing different \rev{guidance methods} is important since different encoding methods may significantly impact \rev{critical factors like} usability and performance.


We conducted an experiment investigating how different design techniques affect motion performance and guidance usability \rev{to inform effective AR design practices 
for various types of motions.
Drawing on existing AR guidance approaches, }
we measure performance and usability across two key design variables (Fig. \ref{fig:visualization}): encoding (path, object, line, human), and perspective (first-person, third-person).
An encoding is the graphical method guiding a person’s motion, such as following a path \cite{tang_physiohome_2015} or copying a human \cite{anderson_youmove_2013}. 
Perspective refers to the point of view from which the person sees the encoding. 
Both perspective and encoding 
\rev{vary across existing guidance systems and may} influence effective motion guidance.


\rev{Inspired by collaborations with physical therapists, we tested these design variables using isolation exercises common in physical therapy (PT).}
Isolation exercises are movements that target specific muscle groups, operate in a single plane of motion, and are restricted to a single joint. These exercises are particularly valuable in \rev{PT} and fitness broadly as they allow patients to work specific muscles, helping in injury recovery, slowing the progression of chronic diseases, and improving general health \cite{reiman_integration_2011, ruegsegger_health_2018}. However, 
\rev{these exercises may be difficult to execute correctly}
without guidance from a physical therapist or personal trainer as they 
require precise movement patterns to properly strengthen intended muscle groups. 
Isolation exercises are an important space for AR motion guidance due to the real-world need for unsupervised exercise guidance and the practical design challenges they raise: these movements 
encompass diverse motion patterns across upper and lower body segments in multiple planes of motion. We collaborated with \rev{PT} clinicians to select three common isolation exercises: 
trunk rotation, shoulder flexion, and hip abduction. 
These exercises 
test methods of guiding movement across various planes of motion, targeting both the upper and lower body and addressing balance, strength, and flexibility goals. 

These movements also vary in 
\rev{\textit{visibility}---
how well people can naturally see their own body 
motion---providing visual feedback to supplement proprioceptive cues.} 
High-visibility movements \rev{(e.g., trunk rotation in Fig. \ref{exercises})}
occur within the natural field of view, while moderate-visibility movements
require people to adjust their gaze and/or head position during the movement to maintain visibility, such as looking down to observe the motion\rev{, which may compromise the movement}. Low-visibility movements like planks or squats 
occur when directly observing one's own body movements is extremely challenging or impossible. 
Movements in the moderate and low visibility range introduce practical challenges for designers, as shifting one's head may either disrupt the quality of movement (e.g., by shifting balance or compromising form) or can be complicated by the use of head-mounted AR displays. However, we lack empirical guidance on how these kinds of motions might affect design. Given the additional design considerations required for 
low-visibility conditions, we focus on high and moderate visibility movements.

\rev{We conducted a 4$\times$2 within-participants experiment to understand the effect of visualization on motion guidance. } 
Key findings from our experiment include: (1) third-person perspectives enhance usability for moderate visibility movements,
(2) line encodings promote 
less precise movements compared to object encodings, and (3) path visualizations increase completion time but do not improve accuracy despite providing more visual cues.


\vspace{3pt}\noindent\textbf{Contributions:} Our contributions include:
 1) an experiment investigating the effects of encoding and perspective on accuracy and usability for guiding visible movements with AR,
 2) a discussion of how different visualization techniques can better support a range of motions, including upper/lower body and different planes of motion, and
 3) empirically-grounded design recommendations and considerations for 
 AR motion guidance.

%% file: 2_Related_Work.tex
\section{Related Work}

\subsection{Mixed Reality Motion Guidance}

Mixed reality (MR)
is increasingly common for
guiding body motion across various domains, including sports training for basketball \cite{lin_towards_2021}, cycling \cite{kaplan_towards_2018}, tai chi \cite{han_my_2017}, martial arts \cite{hoang_onebody_2016}, yoga \cite{jo_flowar_2023}, and dance \cite{laattala_wave_2024-1, anderson_youmove_2013}; 
medical training for surgical procedures and needle injection \cite{heinrich_comparison_2019}; 
assembly and maintenance tasks \cite{gemito_mixed_2023}; 
and guiding hand gestures \cite{wang_exploring_2024, freeman_shadowguides_2009}. These systems employ situated visualizations relative to the user or a digital representation of the user, providing 
spatially-relevant feedback. 
This augmented feedback improves both accuracy and user engagement over traditional video demonstrations \cite{tang_physiohome_2015, anderson_youmove_2013, hoang_onebody_2016}, 
highlighting the potential of MR
to improve exercise effectiveness, 
promote adherence to movement protocols, and improve motion quality and outcomes.

Augmented reality (AR) motion guidance supports both precise gestures and larger full-body movements. 
For example, previous work on hand gestures guides movement
using situated shapes such as lines, circles, and arrows to indicate target and direction \cite{wang_exploring_2024, sodhi_lightguide_2012, delamare_designing_2016}. 
For full-body movement sequences like dance instruction, systems 
focus on visualizations showing the full body, such as AR mirrors and human videos or models  \cite{laattala_wave_2024-1, jo_flowar_2023, yang_implementation_2002, anderson_youmove_2013}. 
Small scale movements tend to emphasize precise feedback, while large scale movements provide more coarse feedback. Many 
movements, such as isolation exercises, require broader movements than hand gestures but 
more precision and focus than prior full-body movement types like dance sequences, requiring further investigation to understand effective design practices. 

Systems can use AR to guide motion with both first-person (1P) and third-person (3P) perspectives. \rev{
Perspective may affect several usability aspects, including embodiment \cite{galvan2017characterizing}, cognitive load \cite{huang2022immersivepov}, and body awareness \cite{debarba2015characterizing}, with the optimal perspective 
likely depending on user 
positions or operating space 
\cite{debarba2015characterizing}.} Salamin et al. found that 3P facilitated better prediction of mobile object trajectories, while 1P 
better supported intricate hand manipulations \cite{salamin_benefits_2006}. This suggests that 3P improves large-scale visibility, whereas 1P enhances 
visible, proximal movements. \rev{However, Medeiros et al.~\cite{medeiros2018keep} found 1P was generally preferable for full-body navigation. }
For martial arts poses, Hoang et al. \rev{also } observed improved accuracy and preference with 1P for upper body static poses \cite{hoang_onebody_2016}. Yu et al. also found that 1P perspective performed best for accuracy and time for visible arm movements \cite{yu_perspective_2020}. However, these studies focus on upper body movements in the frontal plane and may not generalize to more diverse 
exercises, such as 
lower body movements or different planes of motion. 

\subsection{Mixed Reality Rehabilitation Guidance}

Our experiment is directly motivated by our collaborations with physical therapists who want to apply AR approaches to improve remote therapies. 
AR and VR 
have shown promise in medical rehabilitation, with applications in stroke recovery \cite{colomer_effect_2016}, injury rehabilitation \cite{tang_physiohome_2015, ayoade_novel_2014}, and disorders affecting balance \cite{lee_effects_2017}. Several studies have explored the use of AR mirrors to situate target shapes such as lines and wedges relative to participants' reflections for both upper and lower limb \rev{PT} \cite{tang_physiohome_2015, doyle_base_2010, ayoade_novel_2014}. While these systems demonstrated positive outcomes using 3P AR mirrors, other research for non-medical applications suggests that AR mirrors are less preferred and less accurate than 1P interactions in a headset \cite{yu_perspective_2020, hoang_onebody_2016}. 

Methods for cueing and augmenting motion for AR vary. Butz et al. 
reviewed AR applications in medical rehabilitation, identifying key themes and patterns across the research landscape \cite{butz_taxonomy_2022}. Their analysis of visual guidance methods revealed popular techniques for guiding interaction. Targets show a spatial destination to be reached \cite{powell_openbutterfly_2020, doyle_base_2010}. Paths show the movement trajectory \cite{tang_physiohome_2015,vieira_augmented_2015}. Demonstrated interactions use overlays \cite{anderson_youmove_2013} or recordings \cite{jo_flowar_2023} to display proper motion. However, Butz et al. also note a lack of comparative studies examining the efficacy of different 
visual guidance approaches in AR for medical rehabilitation. Though few previous works did compare visual elements \cite{colomer_effect_2016, tang_physiohome_2015, sodhi_lightguide_2012}, they focus on specific encodings and use cases that may not generalize to 
other motions. 
Butz et al. also note that the majority of existing research focused on upper limb rehabilitation \cite{tang_physiohome_2015, vieira_augmented_2015, powell_openbutterfly_2020, williams_augmented_2019}, with lower limb studies primarily concentrating on gait and balance \cite{chen_lower_2020,chang_eeg_2019}. \rev{For example, SleeveAR \cite{sousa2016sleevear} and Physio@Home \cite{tang_physiohome_2015} use AR to improve upper body rehabilitation over traditional video guidance.} Given the varied levels of visibility with different 
types of motions, understanding the effects of visual guidance on different body parts and varying levels of visibility is crucial for developing effective 
guidance. 
While our study is motivated by \rev{PT} applications, we 
aim to generate insight into AR motion guidance broadly; these design challenges are not unique to rehabilitation. 
We 
achieve this by conducting an empirical study comparing visualization techniques with different perspectives and encodings over both upper and lower body 
movements across different planes of motion.

%% file: 3_Motivation_and_Hypotheses.tex
\section{Motivation and Hypotheses}

\begin{figure*}[t]
  \centering                   
  \includegraphics[width=\textwidth]{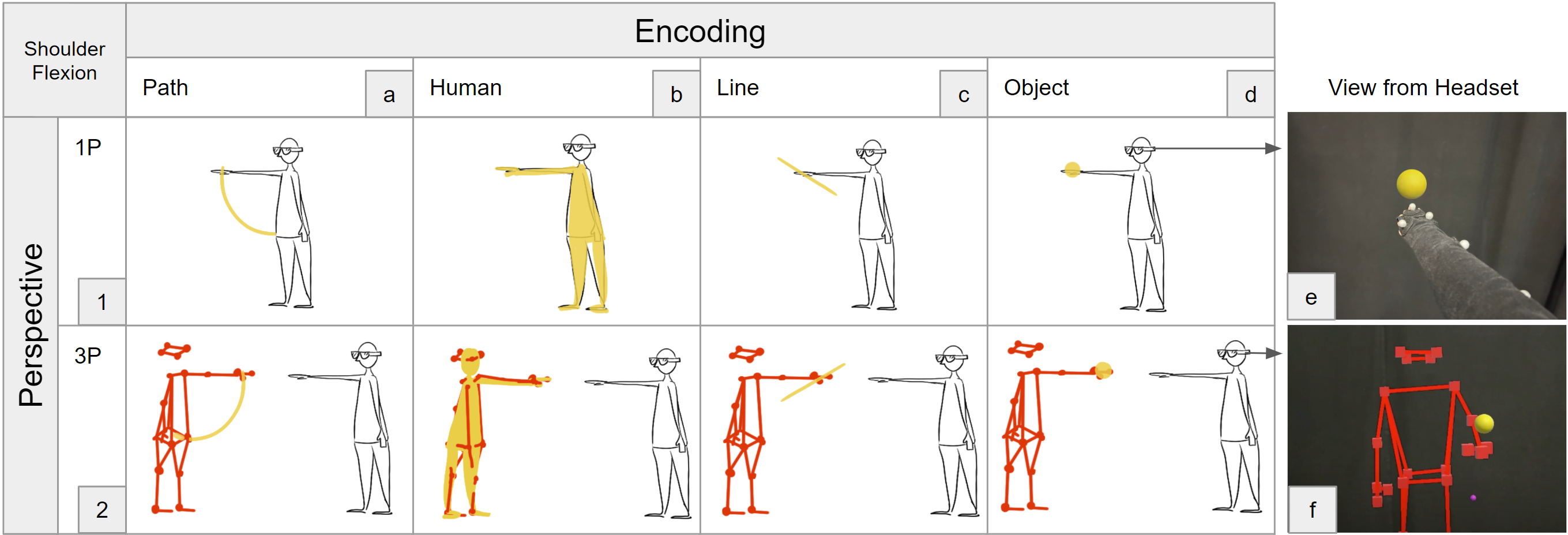}
    \caption{ 
    \rev{Schematic of the tested visualizations for shoulder flexion.} Each visualization has an encoding---a) path, b) object, c) line, and d) human---and perspective---1) first-person and 2) third-person. Examples of object in first person (e) and third person (f) \rev{viewed} from the \rev{HMD}.
    }

  \label{fig:visualization}
  \vspace{-1em}
  
\end{figure*}

Motion guidance is beneficial for rehabilitation, sports training, and careers involving physical tasks such as assembly, building, and surgery. 
%
To ground our investigation, we talked with \rev{PT clinicians}, 
\rev{PT} patients, and collegiate athletes 
to identify key challenges in unsupervised exercise performance that may be addressed with AR motion guidance. 

\rev{PT} is critical in both injury recovery and chronic disease management, 
using prescribed exercise protocols to enhance strength and mobility outcomes 
\cite{shahid_comprehensive_2023, ruegsegger_health_2018}. Patients typically go to in-clinic sessions where physical therapists provide exercise instruction and are prescribed exercises to complete at home \cite{merolli_use_2022}. 
 Many rehabilitation protocols focus on \textit{isolation exercises}: movements that focus on precise, controlled motions using a target joint or muscle group \cite{reiman_integration_2011}. Isolation exercises involve a \textit{target body part}, such as a hand, moving from a start position to a specific \textit{target}.
 The \textit{target path} is the correct trajectory the target body part must follow from the start position to the target to correctly complete the motion. These exercises can target weak or injured body parts and correct muscular imbalances, and are sufficiently simple for individuals at any fitness level to perform effectively. 
Despite being crucial for recovery, patients often find it challenging to perform exercises at home without the clinician's feedback, 
may struggle to know if they are doing the exercise correctly, and worry they may injure themselves \cite{kandel_understanding_2024}. Many isolation exercises prescribed for \rev{PT} are also used for athletic training and general fitness. While athletes participate in structured team practices, they  also independently conduct supplementary strength training targeting sport-specific muscle groups. However, athletes and casual gym goers alike can unknowingly execute exercises with improper form, leading to 
suboptimal performance and potential injury. Given their use across rehabilitation and training, 
isolation exercises 
offer a useful testbed for 
AR-guided movement instruction.
We focus on isolation exercises with a single target body part and corresponding target position as 
these movements enable more precise 
performance measures and represent a majority of \rev{PT} and fitness training practices \cite{reiman_integration_2011}. 

While isolation exercises are used in both rehabilitation and fitness, different populations may have different 
goals.  
For example, elderly \rev{PT} patients 
often aim to maintain regular movement patterns, 
so a therapist might emphasize treatment adherence over precise form or exercise intensity. In contrast, when working with competitive athletes,
coaches are focused on movement performance and proper biomechanics to prevent injury and optimize strength gains. Understanding how visual motion guidance techniques impact performance and usability can enable optimized and personalized systems based on these varying needs. 


We examine two critical design choices for guiding motion in AR---\textit{encoding} and \textit{perspective}--- 
across three 
\rev{common isolation} exercises (see Section \ref{exercises}). For each exercise, we investigated 
combinations of four encodings (\S \ref{encoding}) and two perspectives (\S \ref{perspective}), \rev{as shown in Figure \ref{fig:visualization}}. We selected these designs based on prior 
approaches 
and iterative testing in consultation with clinicians. 

\subsection{Encoding}

\label{encoding}
Our experiment focused on four encodings: path, object, line, and human. While other 
encodings exist, we chose encodings that are well-represented in previous systems \cite{butz_taxonomy_2022}, represent distinct shapes and design affordances, and align with conventional in-person 
motion guidance methods. Rather than teaching movements, we selected encodings effective for \textit{guidance}, which, following conventional PT and athletic training workflows, assumes prior demonstration from a person and requires less detailed instruction in AR.  

\textbf{Path}: The path encoding shows a 3D arc 
indicating the trajectory the target body part should follow (\rev{Figure \ref{fig:teaser},} Figure \ref{fig:visualization}a). The target is located at the end of the path. Previous systems have used both arcs and wedges to guide upper body limb rehabilitation \cite{tang_physiohome_2015, vieira_augmented_2015, ayoade_novel_2014}. We chose a 3D arc instead of an area wedge because the arc 
minimized body occlusion and proved easier to follow in piloting. 

\textbf{Object}: The object encoding presents a sphere at the target location (\rev{Figure \ref{fig:teaser},} Figure \ref{fig:visualization}b), focusing the participant's attention solely on the target \cite{powell_openbutterfly_2020, williams_augmented_2019, bouteraa_training_2019, subramanian_exertiles_2021}. This encoding mirrors common clinical practices where physical therapists ask patients to touch targets at various spatial positions, such as a spot on a wall or the clinician's hand. We render the target as a sphere 
such that the object's visual hull uniformly encompasses the target.

\textbf{Line}: The line encoding displays a long line with the target at its center (\rev{Figure \ref{fig:teaser},} Figure \ref{fig:visualization}c), using the metaphor of a ``finish line'' \cite{doyle_base_2010}. For each exercise, we oriented the line orthogonal to the target path to ensure participants would pass through the line's center when completing the exercise correctly.

\textbf{Human}: The human encoding shows a full-body human mesh in the final motion pose (\rev{Figure \ref{fig:teaser},} Figure \ref{fig:visualization}d). Previous applications have used virtual humans to guide full body sequences (\cite{laattala_wave_2024-1, anderson_youmove_2013, hoang_onebody_2016}). This encoding models in-clinic practices where physical therapists demonstrate movements or have patients look in the mirror to see their own body. We used a mono-colored, gender-neutral human mesh
rather than a simplified skeletal representation as 
the mesh model offered less ambiguity in distinguishing joints and more motion clarity 
when dynamic movement cues are absent.


\subsection{Perspective}
\label{perspective}


\textbf{First-person}: 1P renders the encoding 
on the participant’s physical body (\rev{Figure \ref{fig:teaser},} Figure \ref{fig:visualization}.1). The participant completes a motion by directly aligning their own 
target body part with the virtual target. 1P perspectives may improve engagement and performance 
due to the added sense of presence and the intuitive spatial alignment at true-to-life scale \rev{\cite{debarba2015characterizing, yu_perspective_2020, hoang_onebody_2016}}. However, visualizations positioned closer to the user risk being partially obscured due to the 
headset's 
field of view \cite{ren_evaluating_2016}. 
Lower visibility 
motions may require shifting gaze or head position to see, potentially compromising 
motion form or participant comfort and balance. 

\textbf{Third-person}: 3P superimposes the encoding onto a digital representation (i.e., a digital twin \cite{botin-sanabria_digital_2022}) of the person \cite{anderson_youmove_2013, tang_physiohome_2015} (\rev{Figure \ref{fig:teaser},} Figure \ref{fig:visualization}.2). The participant looks forward to see their digital twin facing them and mirroring their real-time movement,  similar to looking in a mirror as common in PT, and aligns the twin's corresponding body part with each target. \rev{However, unlike a mirror, 
participants in our study see only an abstract skeletal twin (e.g., Fig. \ref{fig:teaser}, bottom) rather than their actual body, with depth cues provided through 3D rendering instead of 2D reflection}. 
We use a red skeleton digital twin to distinguish it from the human encoding
as the real-time movements made it easy for people to distinguish joints in the skeleton in piloting. We piloted side-views of the digital twin for motions in the sagittal plane (Figure \ref{fig:poses}.2), but people preferred the familiarity of a mirror-like front-view. 
Though 3P does not provide the same sense of presence and physical body alignment as 1P \rev{\cite{debarba2015characterizing,galvan2017characterizing}}, 
3P allows individuals to simultaneously observe the entire body \rev{via the twin skeleton} and the full visualization while maintaining a forward gaze.

\subsection{Hypotheses}
\label{hypotheses}
Drawing on past work and experiences, we hypothesized:

\textit{H1: Path encoding will have the highest accuracy.} 
The path's continuous visual guidance 
may enhance participants' understanding of the tested motions and facilitate accurate real-time adjustments \cite{vieira_augmented_2015}. Other encodings lacking trajectory information may lead participants to take incorrect paths or misalign with the target due to insufficient contextual cues. 

\textit{H2: 1P will be more accurate for higher visibility movements, while 3P will be more accurate for lower visibility.} Through direct, real world physical alignment, 1P provides better accuracy for upper limb instruction compared to 3P \cite{yu_perspective_2020, hoang_onebody_2016}. However, for exercises that require looking down to see the motion, 1P alignment becomes more challenging due to head tilts and potential body occlusions disrupting visibility of the visualization \cite{ren_evaluating_2016}. 3P may mitigate these issues by allowing people to maintain a neutral head position recommended by physical therapists for most exercises, as well as 
minimizing body occlusions and offering a more natural vantage point for observing full-body movements.

\textit{H3: Object encodings will be perceived as more usable.} The clear, singular focus point of a target object allows people to concentrate on the the target without processing additional visual information. Physical therapists 
use tangible targets on walls, shelves, and floors to guide patients through reaching and balance exercises, enhancing engagement and precision \cite{subramanian_exertiles_2021}. Other encodings contain additional graphical elements, which potentially increase the mental effort required for interpretation and interaction \cite{tang_physiohome_2015}.

\textit{H4: 1P will be more usable for higher visibility movements, whereas 3P will be perceived as more usable for lower visibility.
} Compared to 1P, 3P requires more 
mental transformation to match body motion to the corresponding digital twin motion, which may make  tasks harder \cite{yu_perspective_2020}. However, for reduced visibility exercises,
participants 
may find looking down with 1P unnatural and uncomfortable. In these scenarios, we expect participants to prefer 3P as it allows for a neutral head and neck position, more visibility of the body and visualization, and improved motion flow (i.e., uninterrupted, continuous movements) \cite{jo_flowar_2023}.

%% file: 4_Methods.tex
\section{Methods}

\subsection{Stimuli}
We conducted a 4 (encoding) x 2 (perspective) within-subjects experiment to investigate the effects of encoding and perspective on performance and usability across three movements. 
Participants saw 
24 total visualizations built in Unity and deployed in a HoloLens 2 headset. Participants performed study exercises in a motion capture space wearing a suit with 42 markers 
placed according to the Vicon Human Body Pose Full Gait Template protocol.\footnote{https://help.vicon.com/space/Nexus215/11376865/Full+body+modeling
+with+Plug-in+Gait} The tracking system consisted of 9 Vicon Valkyrie VK8 cameras, which were used to capture body motion coordinates for rendering the digital twin 
and calculate target and target path locations. Given that the goal of our research was to 
understand motion guidance broadly, we prioritized choosing a motion capture system with higher accuracy rather than the in-home systems common to our motivating use cases. 

For 1P visualizations, participants saw each visualization \textit{in situ}, scaled to their body position and size. For 3P, participants saw each visualization \rev{on} a digital twin skeleton rendered 2m in front of them using red 2cm wide lines connecting 2cm cubes at major joints. 
Visualizations consisted of four encoding types: path, object, line, and human guide (\S \ref{encoding}). 
The path encoding used 2cm wide 2D lines
to trace a continuous path from the start position to the end position based on optimal arcs demonstrated by clinicians and scaled to the 
participant's limbs. The object encoding used a 8cm radius sphere, centered at the target coordinate (see \S \ref{procedure}). 
The line encoding was a 1m long $\times$ 2cm wide 2D horizontal line centered at the target coordinate. 
The human encoding used a SMPL gender neutral human mesh \cite{loper_smpl_2015} demonstrating the exercise's final target pose.
The SMPL model was manually scaled and positioned based on the participant's body size with the target body part aligned with the target.

 We selected colors that 
 \rev{had high contrast to the environmental background and each other and were robust to color vision deficiencies}: yellow and blue for the targets and red for the digital twin. When the target body part was within
 8cm of the target position, the encoding changed colors from yellow  to blue  to indicate success. The 1P human was partially transparent so that participants could better see their own body and the human visualization at the same time. All other graphical encodings 
 were rendered at the full opacity enabled by the Hololens 2.  
We rendered the 3P human as fully opaque to avoid confusion regarding the positioning of the 3P human body parts, which emerged in piloting. 
\rev{These 
values were determined through internal pilot testing, 
and prior literature, 
 prioritizing visibility (
contrast and color accessibility), interpretability (
alignment with existing semantics), and HoloLens 2 feasibility
\cite{tang_physiohome_2015, hoang_onebody_2016-1, doyle_base_2010-1, williams_augmented_2019}.}
All infrastructure and data is available at \url{<https://osf.io/4nhtf/?view_only=b5168623250a44aa826ab027df42d87b>}

\subsection{Tasks}
\label{exercises}
\begin{figure}[t]
  \includegraphics[width=\columnwidth]{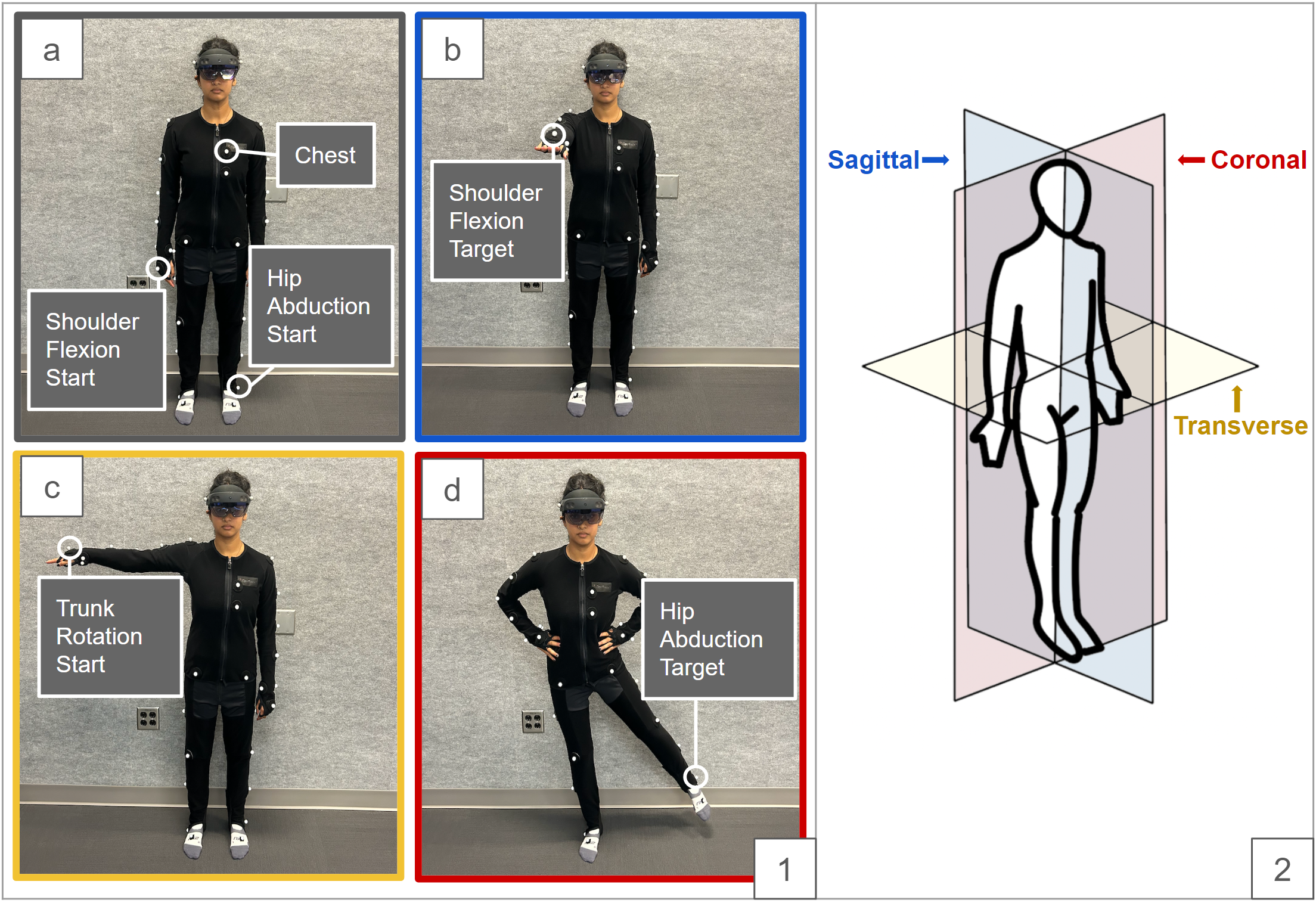}
  \caption{(1) At study start, participants held four poses to capture coordinates for shoulder flexion (b), trunk rotation (c), and hip abduction (d). These exercises occur in different cardinal planes (2): sagittal, transverse, and coronal.
  }
  \label{fig:poses}
  \vspace{-1em}
\end{figure}

Participants were asked to physically move a target body part from a specified starting point to a target location. In consultation with clinicians, we determined target body parts and target paths using three 
common isolation exercises: trunk rotation, shoulder flexion, and hip abduction. 
\rev{We selected exercises 
to include upper and lower body exercises, movement in all three planes of motion (Figure \ref{fig:poses}.2),
and varying levels of visibility, which may impact visualization usability and performance.}
To 
encourage generalizability, we focused on 
isolation exercises that could be performed safely and effectively regardless of age or fitness 
that are used in a range of activities across PT and fitness \rev{both independently and as components of more complex movements}.


\textbf{Trunk Rotation}
focuses on the person's back and abdomen, increasing flexibility in the trunk. The person starts with their feet shoulder width apart, facing forward, with their right arm out to the side (Figure \ref{fig:poses}.1c). With their feet facing forward, the person turns their torso to the left, making an arc with their right hand \rev{parallel to the ground} in the transverse plane of motion until their hand is on the other side of their body (Figure \ref{fig:teaser}). Trunk rotation has high visibility because the target body part remains in the person's natural field of view throughout the entire range of motion.

\textbf{Shoulder Flexion}
focuses on strengthening shoulder musculature (e.g., anterior deltoid, bicep). The person starts with their arm down by their side \rev{aligned with their hip}, and then while keeping their arm straight, lifts their arm up \rev{perpendicular to the ground} in the sagittal plane until 
\rev{it reaches shoulder level} (Figure \ref{fig:poses}.1b). Shoulder flexion has moderate visibility as people must 
look downward and to the side to see the initial phase of the movement.

\textbf{Hip Abduction}
\rev{starts with the feet 
shoulder-width apart on the floor. Participants move} one leg away from the body \rev{towards their} side in the coronal plane of motion (Figure \ref{fig:poses}.1d). 
Variations of the hip abduction exercise 
serve different therapeutic purposes. For example, using a table or chair for support, the person could complete this action to target leg strength or hip flexibility. Since trunk rotation works on flexibility and shoulder flexion works on strength, we chose to remove support to challenge balance. 
Hip abduction has limited visibility as 
people must look downward to see the leg movement, which occurs entirely below waist level.

\begin{figure*}[t]
  \vspace{-1em}
  \centering                   
  \includegraphics[width=\textwidth]{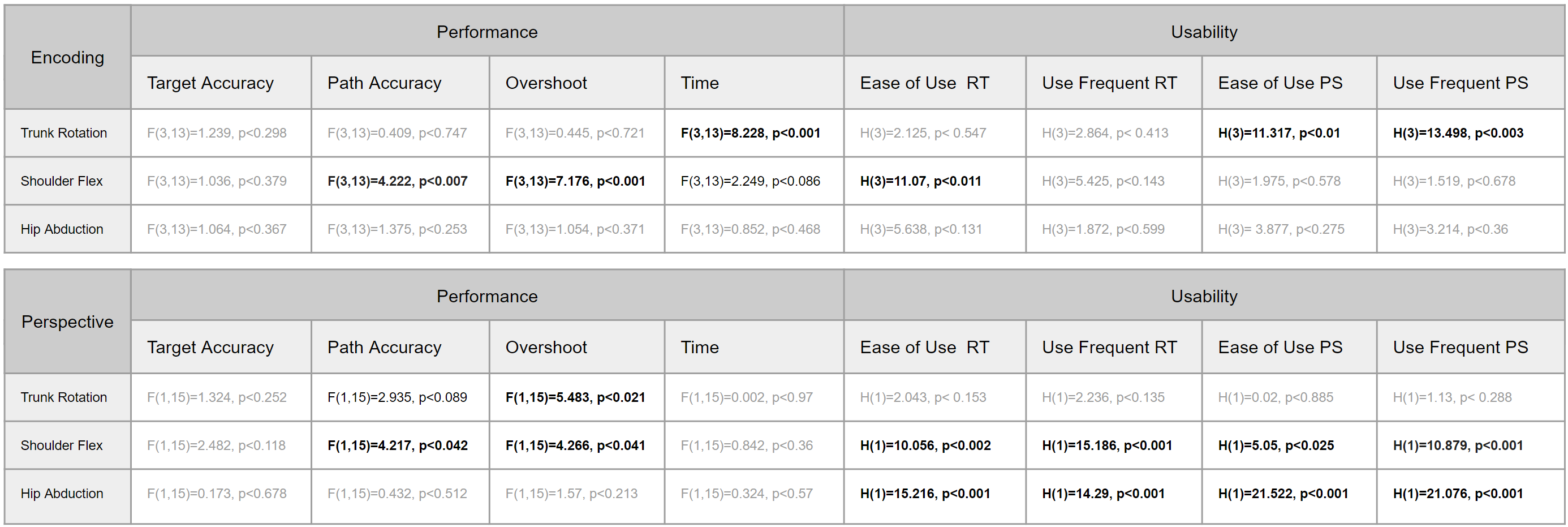}
    \caption{ Results from our ANOVA and Kruskal-Wallis tests. Bolded text indicates significant findings ($p<0.05$), standard text indicates marginal effects ($p<0.1$), and grey text indicates non-significant effects.
    }
  \label{fig:table}
  \vspace{-1em}
\end{figure*}

\subsection{Procedure}
\label{procedure}

Participants provided informed consent in accordance with our IRB protocol, completed a demographics survey, and were fitted with the motion capture suit and HoloLens 2. 
Participants then completed 
three calibration procedures: Vicon calibration, coordinate space calibration, and task calibration. 

For Vicon calibration, participants assumed a T-pose to enable accurate body joint labeling within the Vicon Nexus software. We used the Vicon Datastream SDK to stream marker labels and coordinates to the Unity environment in real-time. 

For coordinate space calibration, we implemented a sampling-based approach to establish a linear transformation between the HoloLens 2 and Vicon coordinate systems. Participants walked in a circular path while 1,000 paired coordinates (markers on the HoloLens 2 headset and Unity camera) were collected. We used the Kabsch algorithm \cite{kabsch_solution_1976} to derive the transformation matrix between these two point sets, with a root mean squared deviation $< 0.04$. This process ensured accurate AR overlay positioning relative to the participant's body and headset orientation. 

For the task calibration, we recorded key participant body marker coordinates to compute individual body proportions used to render the visualizations. 
Based on clinician feedback, we modeled the trunk rotation target path as a circular arc and the shoulder flexion and hip abduction as elliptical curves. Participants stood on a marked line and assumed four poses (Figure \ref{fig:poses}.1) to capture relevant coordinates. For shoulder flexion and hip abduction, elliptical paths were computed 
as $\frac{x^2}{a^2} + \frac{y^2}{b^2} = 1$, where start (Figure \ref{fig:poses}.1a) and target (Figure \ref{fig:poses}.1b and Figure \ref{fig:poses}.1c) positions defined the ellipse parameters. The trunk rotation path was generated by rotating the start position (Figure \ref{fig:poses}.1c) 180° around the participant's chest coordinate (Figure \ref{fig:poses}.1a). To ensure proper visualization alignment, 
the participant's forward vector was recorded using Unity's camera orientation during the neutral standing position (Figure \ref{fig:poses}.1a). This vector served as a reference for situating all subsequent visualizations. To align with the visualizations, participants stood on the marked line and faced forward for each trial as in calibration. 

Following 
calibration, participants were 
given an overview of the study objectives and experimental protocol. The researcher then conducted a standardized demonstration of the three exercises emphasizing correct foot position and  body posture. We designed this instruction phase to minimize the potential for movement artifacts from task misunderstanding, ensure consistency across participants, and reflect 
real-world practices where 
people learn 
movement sequences with an instructor before practicing independently, 
reflecting our 
goal of guiding people through known movements rather than teaching new exercises. 

Participants completed 24 trials consisting of 8 visualization conditions (4 encodings $\times$ 2 perspectives) for each of the 3 exercises: trunk rotation, shoulder flexion, and hip abduction. To ensure consistent range of motion, we marked the starting position of each exercise with a small pink sphere \rev{(Fig. \ref{fig:teaser}). To 
reduce muscle memory effects, we shifted the final target 
along the correct motion path 
randomly from 0--20cm.} For each trial, participants first had two attempts to familiarize themselves with the motion and encoding. 
They then completed three cycles of each exercise for the presented 
visualization. Multiple cycles allowed for a more robust understanding of the participant's performance by sampling over individual performance variability. 
Following the completion of the three cycles, participants 
could provide verbal feedback and responded to two Likert scale questions 
used in prior usability studies 
\cite{vlachogianni_perceived_2022}: ``On a scale of 1--5, did you find the visualization was easy to use?'' and ``On a scale of 1--5, would you like to use this system frequently?''. 
To control for potential transfer effects, 
we randomized visualization order for each participant. The order of exercises remained constant (trunk rotation, shoulder flexion, hip abduction) mirroring typical 
protocols where 
people complete exercises in a prescribed order. As participants completed the trials, the marker coordinates of the target body part were recorded at a sampling rate of 30 Hz to collect motion data. 

After completing the formal trials, participants 
ranked the eight visualizations based on their ease of use and anticipated frequency of use, 
giving
unique rankings with no ties. 
Each question showed sketches of 
all eight visualizations to aid in recall. We compensated participants with a \$20 Amazon gift card.

\subsection{Measures}
We measured both performance and usability for each visualization. 
Performance was evaluated using five primary metrics: 1) path error, quantified using normalized Dynamic Time Warping (DTW) to measure the degree of alignment between the participant's movement trajectory and the target path; 2) target error, calculated as the minimum Euclidean distance between the target position and the participant's target body part end position; 3) completion time, defined as the duration from initial movement from the specified starting position to the target position;
and 4) overshoot frequency, defined as the number of instances where the participant's target body part passed the intended target position. 
We computed the mean completion time, path errors, and target errors across the three repetitions for each trial to derive a single value for each measure per participant per trial. The error metrics and overshoot 
reflect controlled and proper form, while 
time indicates performance speed.

We evaluated usability through 
1) two Likert scale questions administered immediately following each trial to capture real-time feedback on individual visualization techniques and 2) comparative rankings collected post-study. This combination of immediate and retrospective feedback 
characterized both isolated and comparative user experiences.
We additionally collected open-ended verbal feedback after each trial and at the end of the study. 

\subsection{Participants}
We recruited 18 participants (11 female, 7 male) from the local campus community, \rev{
consistent with prior work in AR motion guidance \cite{tang_physiohome_2015, sodhi_lightguide_2012, anderson_youmove_2013, hoang_onebody_2016} and local standards in HCI \cite{caine2016local}.} Participant ages ranged from 18--49 ($\mu = 23.44, \sigma = 7.49 $). We excluded one participant's performance results due to data recording errors but include their usability data in our analysis. Participants self-reported on a 5-point Likert scale 
moderate previous exposure to \rev{PT} ($\mu = 2.56, \sigma = 1.38$), moderate athletic ability ($\mu = 2.83, \sigma = 0.92$), and moderate to low prior experience with AR ($\mu = 2.28, \sigma = 1.38$). This moderate familiarity with both AR and exercise allowed for 
feedback without the expert biases that might overlook fundamental usability challenges faced by novice users. 
\rev{We recruited healthy young adults 
based on our PT collaborators' guidance to first test a broader range of designs with healthy 
people before moving to more vulnerable populations at higher risk for falls. 
Future work should extend our results to 
more diverse populations in specialized studies ( \S\ref{limitations}). }

%% file: 5_Results.tex
\section{Results}

\begin{figure*}[t]
  \centering                   
  \includegraphics[width=\textwidth]{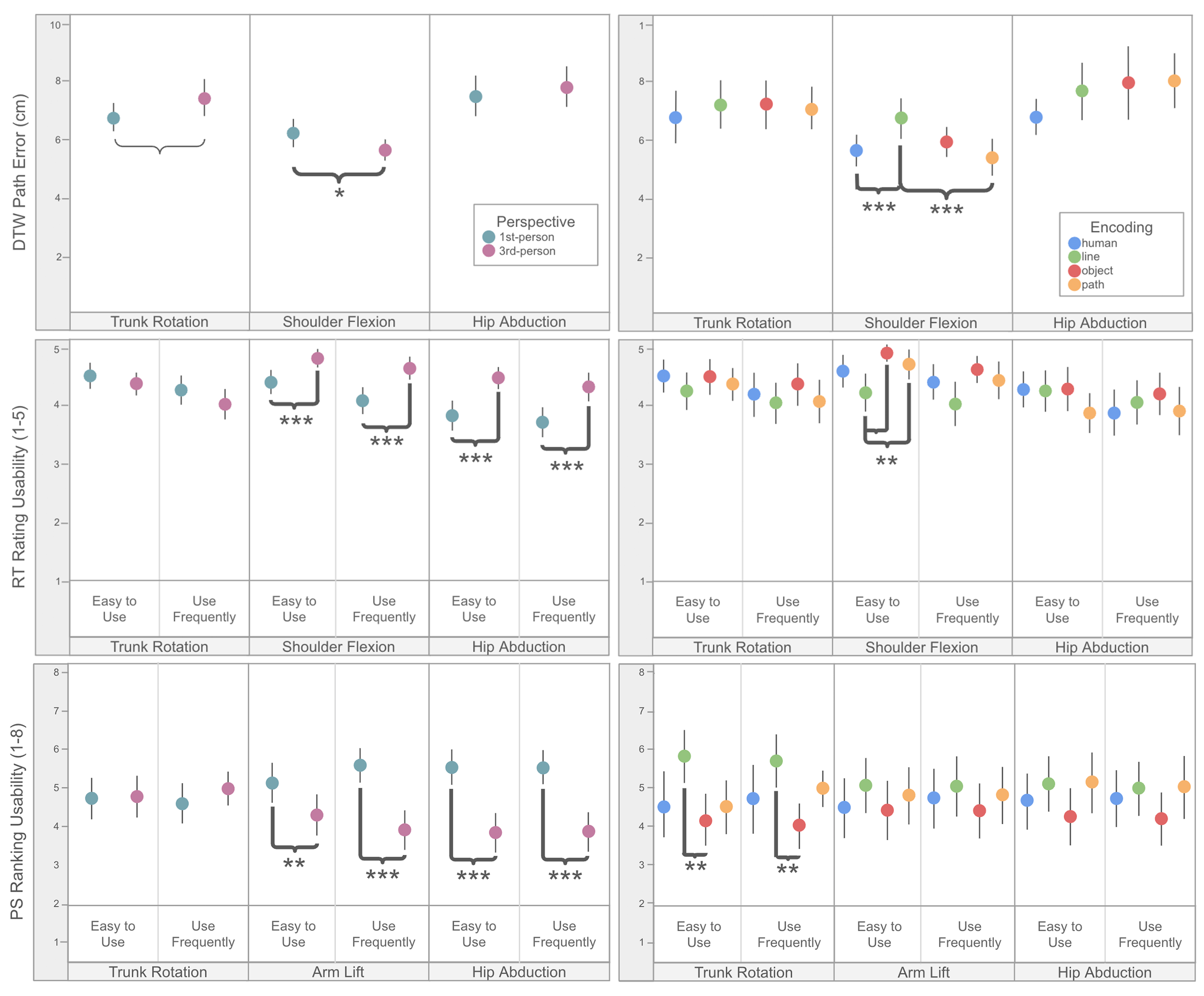}
    \caption{Means and 
    $95\%$ confidence intervals for DTW path error, real-time usability ratings 
    (1--5, where 5 is the most usable), and post-survey usability rankings 
    (1-8, where 1 is the most usable). 
    \textbf{*} indicates significance with $p < 0.05$, \textbf{**} indicates $p<0.01$, and \textbf{***} indicates $p<0.001$.
    }
  \label{fig:plot}
  \vspace{-1em}
\end{figure*}

\begin{figure*}[t]
  \includegraphics[width=\textwidth]{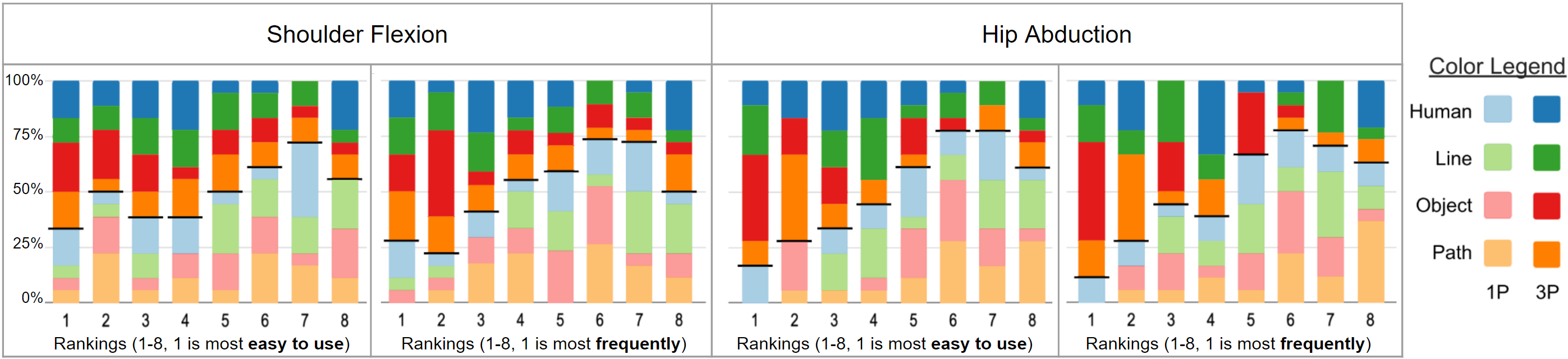}
  \caption{Ranking comparison of visualization techniques for shoulder flexion and hip abduction exercises by ease of use and anticipated frequency (lower ranking = higher usability). First-person (1P, darker colors) and third-person (3P, lighter colors) perspectives are shown for human, line, object, and path visualizations. 3P and object visualizations consistently received better usability rankings.
  }
    \label{fig:stackedbarchart}
    \vspace{-1em}
\end{figure*}

We analyzed performance using full-factorial two-way ANOVAs for each exercise, with the path error, target error, time, and overshoot rate as dependent variables and encoding and perspective as independent variables. 
We collected real-time (RT) and post-study (PS) rankings for both ease of use (E, 5-point Likert scale) and 
anticipated frequency of use (F, 8-point ranking). We used the Kruskal-Wallis test to analyze usability, with RT ease of use, PS ease of use, RT use frequency, and PS use frequency responses as dependent variables and encoding and perspective as independent variables. All post-hoc tests used Tukey Honest Significant Difference (HSD) test ($\alpha = 0.05$). Figure \ref{fig:table} summarizes our results. 
For significant findings, we provide means and $95\%$ confidence intervals for transparent statistical communication \cite{dragicevic2016fair}. 

\subsection{Performance}
Participants successfully reached the target 
in most trials (93$\% \pm 7\%$). We found no significant effects of either perspective or encoding methods on target error.
As a result, we focus on the path error, time, and overshoot to understand performance. 
We 
only report primary effects as we found no significant interaction effects. 

\textbf{Encoding:} 
The path encoding generally took longer to complete but did not result in lower path error, failing to support our hypothesis that that path encoding would have the highest accuracy (H1).
For trunk rotation, we observed that the path encoding (8.15s $\pm$ 1.9s) took significantly longer to complete compared to the object (5.57s $\pm$ 0.69s), line (5.2s $\pm$ 0.63s), and human (6.02s $\pm$ 0.82s) encodings. For shoulder flexion, path took marginally longer to complete 
than the other encodings. 

The line encoding 
had higher path error compared to other encodings. For shoulder flexion, participants had significantly lower path error with the path encoding (5.24cm $\pm$ 0.62cm) and human encoding (5.47cm $\pm$ 0.53cm) than the line encoding (6.54cm $\pm$ 0.65cm), shown in Figure \ref{fig:plot}. The line encoding also resulted in a higher overshoot rate ($25\% \pm 13\%$) compared to the path ($4\% \pm 5\%$), object ($2\% \pm 4\%$), and human ($8\% \pm 8\%$) encodings. 

\textbf{Perspective:} 
We found 
that accuracy 
varied across perspectives depending on the exercise and its visibility, 
partially supporting our hypothesis that 1P would be more accurate for higher visibility movements
and 3P for lower visibility (H2).
For trunk rotation, 1P (6.52cm $\pm$ 0.48cm) was marginally more accurate than 3P (7.2cm $\pm$ 0.61cm), whereas, for shoulder flexion, 3P (5.4cm $\pm$ 0.35cm) was significantly more accurate than 1P (6cm $\pm$ 0.464), shown in Figure \ref{fig:plot}.  
3P resulted in significantly more overshoots compared to the 1P perspective for both trunk rotation (3P: 17.6$\%\pm± 7.5\%$ versus 1P: 6$\% \pm 5\%$) and shoulder flexion (3P: 15$\% \pm 7.6\%$ versus 1P: 5$\% \pm 5\%$).
This finding aligns with increased difficulty in precise limb alignment with the target in 3P. 

\subsection{Usability}

\textbf{Encoding:} Participants generally found object encodings more usable than line encodings, partially supporting H3.
For trunk rotation, PS rankings showed that participants found the object encoding ($\mu_E$ = 3.88 $\pm$ 0.7, $\mu_F$ = 3.88 $\pm$ 0.6) more usable than the line encoding ($\mu_E$ = 5.58 $\pm$ 0.7, $\mu_F$ = 5.44 $\pm$ 0.7). For shoulder flexion, participant's RT feedback indicated that  
the object encoding ($\mu_E$ = 4.77 $\pm$ 0.15, $\mu_F$ = 4.49 $\pm$ 0.23) and path encoding ($\mu_E$ = 4.58 $\pm$ 0.24, $\mu_F$ = 4.3 $\pm$ 0.3) were easier to use than the line encoding ($\mu_E$ = 4.11 $\pm$ 0.32, $\mu_F$ = 3.91 $\pm$ 0.36), shown in Figure \ref{fig:plot}. 

In PS hip abduction rankings, participants preferred the 3P object encoding ($\mu_E$ = 2.83 $\pm$ 1, $\mu_E$ = 2.83 $\pm$ 0.9) over the 1P path ($\mu_E$ = 6.11 $\pm$ 0.85, $\mu_F$ = 6.22 $\pm$ 0.9), 1P line ($\mu_E$ = 5.67 $\pm$ 0.9, $\mu_F$ = 5.56 $\pm$ 0.8), and 1P object ($\mu_E$ = 5.06 $\pm$ 0.91, $\mu_F$ = 5.06 $\pm$ 0.87) encodings. Participants also indicated that they would use the 1P path encoding less often than the 3P path ($\mu_E$ = 3.61 $\pm$ 1.13, $\mu_F$ = 3.39 $\pm$ 1.12), 3P line ($\mu_E$ = 4 $\pm$ 1.04, $\mu_F$ = 4 $\pm$ 1.12), and 3P object ($\mu_E$ = 2.83 $\pm$ 1, $\mu_F$ = 2.83 $\pm$ 0.9) encodings. Figure \ref{fig:stackedbarchart} shows that 3P object was commonly ranked as most usable. 

\textbf{Perspective:} 3P was perceived as more usable than 1P for exercises with more limited visibility, supporting 
H4. 
The RT feedback rated 3P as significantly more usable than 1P for shoulder flexion (3P: $\mu_E$ = 4.68 $\pm$ 0.15, $\mu_F$ = 4.51 $\pm$ 0.6 vs 1P: $\mu_E$ = 4.28 $\pm$ 0.19, $\mu_F$ = 3.97 $\pm$ 0.22) and hip abduction (3P: $\mu_E$ = 4.34 $\pm$ 0.19, $\mu_F$ = 4.19 $\pm$ 0.24 vs 1P: $\mu_E$ = 3.72 $\pm$ 0.25, $\mu_F$ = 3.6 $\pm$ 0.25). Similarly, participants' PS rankings for 3P rated 3P visualizations as better on average than 1P for shoulder flexion (3P: $\mu_E$ = 4.07 $\pm$ 0.53, $\mu_F$ = 3.87 $\pm$ 0.56 versus 1P: $\mu_E$ = 4.93 $\pm$ 0.54, $\mu_F$ = 5.152 $\pm$ 0.48) and hip abduction (3P: $\mu_E$ = 3.61 $\pm$ 0.52, $\mu_F$ = 3.63 $\pm$ 0.54 versus 1P: $\mu_E$ =5.38 $\pm$ 0.47, $\mu_F$ = 5.38 $\pm$ 0.46), (Figure \ref{fig:plot} and Figure \ref{fig:stackedbarchart}).

%% file: 6_Discussion.tex
\section{Discussion}
We measured the influence of visualization techniques for guiding 
motion in AR. We found that both encoding and perspective influenced the usability and performance of visualizations, 
and that the most effective designs for cueing motion depend on 
motion visibility \rev{(i.e., how much of one's own body is visible when completing the motion)} and 
user goals (e.g., prioritizing accurate execution or consistent adherence). These findings provide empirical guidance for future AR 
motion guidance approaches.

\begin{figure}[t]
    \centering
  \includegraphics[width=\columnwidth]{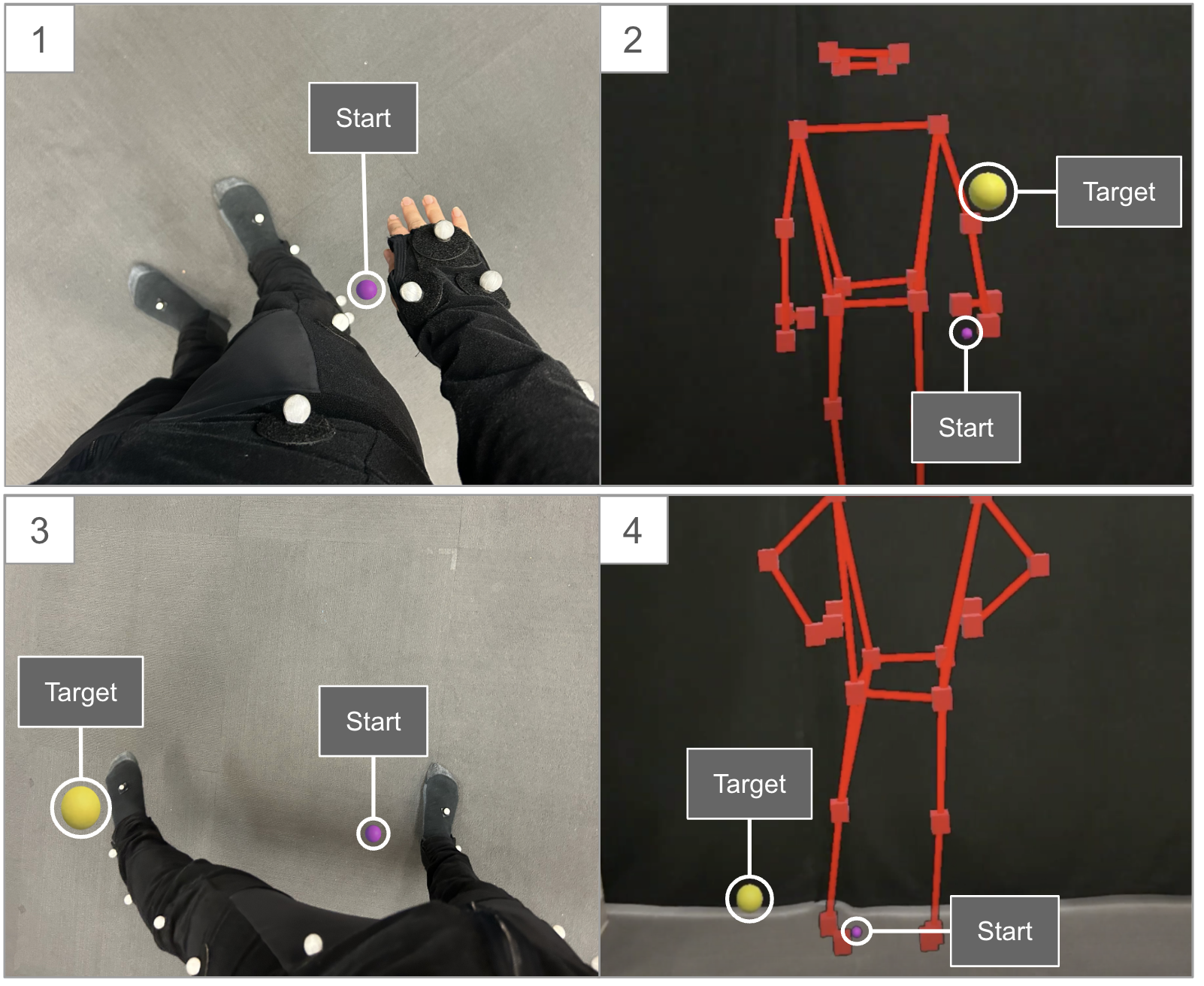}
  \caption{Object encoding visualizations show pink dot (start position) and yellow sphere (target). 1P shoulder flexion (1) and hip abduction (3) require looking down, while 3P versions (2,4) allow neutral head pose. Labels added for clarity, not part of original visualization.
  }
    \label{fig:persp}
    \vspace{-1em}
\end{figure}

\textit{Perspective Depends on Visibility:} While previous studies suggested 1P as superior for both performance and preference for motions with high visibility \cite{hoang_onebody_2016,yu_perspective_2020}, our findings demonstrate that the effectiveness of perspective is heavily dependent on the visibility of the motion being performed. \rev{ Visual body feedback improves movement execution by complementing proprioception \cite{van1999integration} 
.} We found 3P was significantly more usable for movements \rev{where people needed to look down to see their relevant body parts in the 1P view. Specifically, a} majority (83\%) of participants expressed frustration 
when they needed to look down during shoulder flexion or hip abduction exercises in 1P (Figure \ref{fig:persp}), \rev{feeling that the downward gaze was uncomfortable, compromised movement quality, and affected balance.} This discomfort was expressed in comments such as ``when looking down at the visualization, my neck is turning weird'' (P1) and “the more the visualization goes lower, the more annoying it is to look down” (P2). The requirement to look downward also raised concerns about maintaining proper form during exercises. For instance, P11 noted, ``(For shoulder flexion, it’s) hard to see the pink dot at the beginning, I need to crank my neck, so my body is not neutral. For hip abduction, it's hard to balance while looking down at my foot.” P12 expressed similar concerns about posture, ``I don't think I am supposed to look down, looking down throws me off.” Due to these issues with 1P, participants generally preferred 3P, ``I definitely like the red skeleton more because I can look forward'' (P17).

While 3P was preferred for movements requiring a downward gaze in 1P, some participants found 1P more suitable for trunk rotation\rev{---a high-visibility movement completed entirely within the field of view}. When shown the path in 3P trunk rotation, P13 was confused because from the their perspective ``it's not obvious it's a track, it looks like a line.'' P7 and P13 did not like looking forward during the trunk rotation in 3P, because its ``hard trying to face forward to look at the visualization while rotating'' (P7). 
\rev{As in Debarba et al.~\cite{debarba2015characterizing}, the spatial position of targets in a movement heavily influenced preferred perspective.} 
We found that the ideal perspective may vary depending on the specific exercise and its required head and neck positioning. 
Developers should consider the natural gaze direction and head position required for proper form. Choosing 3P for exercises requiring a sustained forward gaze and 1P 
when head movement is part of the intended motion could significantly enhance user comfort and exercise effectiveness.

\rev{For high and moderate visibility movements like shoulder flexion, 3P with see-through HMDs like the Hololens 2 provides an added potential benefit of seeing both one's own body (albeit without added visual cues) and a full digital twin, increasing the available visual feedback. Unlike mirrors or systems using real imagery like Physio@Home \cite{tang_physiohome_2015}, 
we found even abstract twins with AR depth cues enable effective motion cueing, confirming findings on visio-motor alignment in 
3P perspectives \cite{galvan2017characterizing}. Since realism affects task performance and usability across perspectives \cite{debarba2015characterizing,medeiros2018keep}, future work should consider how 
twin representation and 
immersive AR influence cueing perspective.}

\textit{Encoding Affordances for Motion Precision:} People were generally comfortable using all four encodings: on average, each ranked between 4 and 5 on a 5-point Likert Scale for both ease of use and anticipated use frequency. However, people consistently found the line encoding less usable, and it had lower overall performance compared to other encodings, particularly object. People desired a more clearly defined target, as noted by P13: ``It's hard to tell where the middle of the line is. I also can't see the whole line with the headset.'' A shorter line or larger field of view
might have made it easier to find the line's center, but 
people would still need to estimate its position. However, some participants appreciated the analogy of breaking through a line and the broader threshold of success it provided. P3 mentioned, ``I like breaking through the line. It feels like the line is more friendly for doing the exercise, I like the story of not going to a specific position.'' When reflecting on their experience with the line in hip abduction, P8 said, “It’s really hard to get your foot to a very specific point. I would rather have a general area of success rather than a specific point because it feels easier. I like seeing more visual cues with the longer line.” Breaking through the line can also help people push themselves and go past target goals, reflected in the higher overshoot rate for line encodings. However, this overshoot 
may also result in unwanted movement patterns, such as overextending a joint. 

In contrast, people preferred the object encoding's simplicity and clarity as it effectively indicated a specific target point and afforded more precise motions. Participants described it as ``very simple'' (P6) and ``the best one'' (P8). P16 compared the object encoding to the line encoding, saying, ``I like (the object encoding) better than the line because it gives you the exact point to go to.'' Object encodings may be preferable for 
precise positioning, while line encodings accommodate exercises where a general direction of motion is more important than a specific endpoint. 

\begin{figure}[t]
    \centering
  \includegraphics[width=\columnwidth]{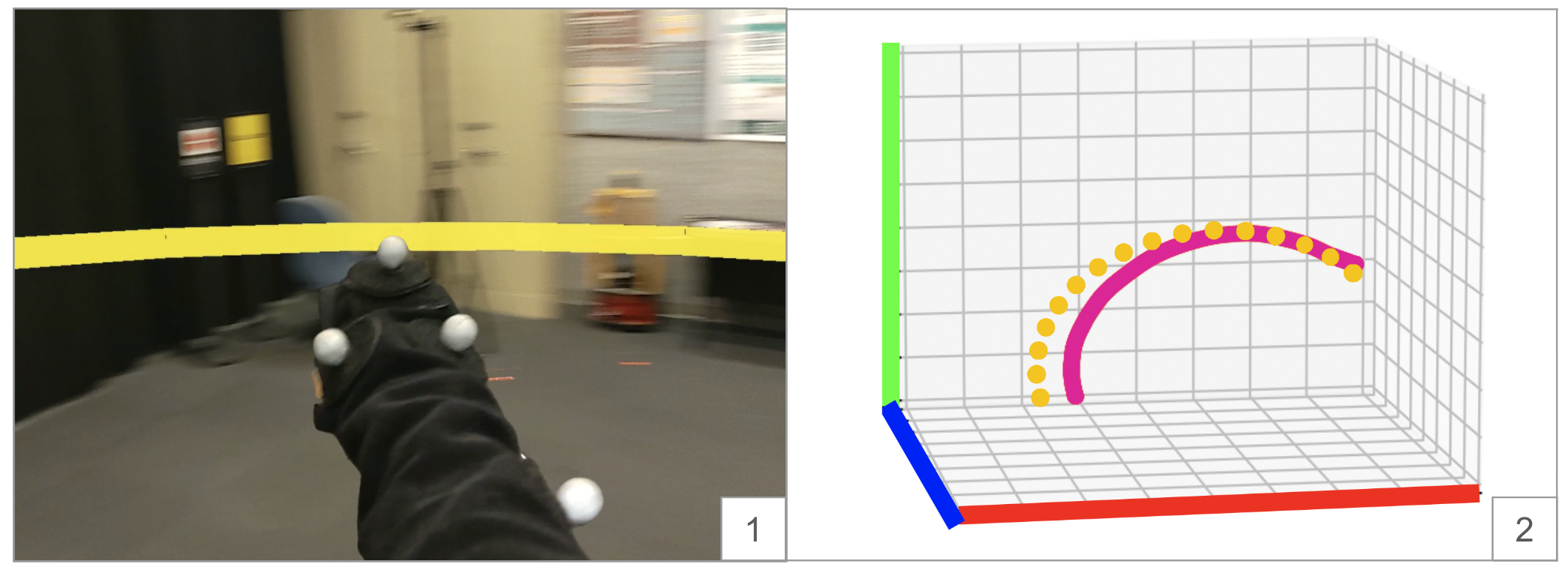}
  \caption{ 1) Birds-eye view showing 3D hand trajectory (pink) versus path encoding (yellow), with x-axis (red), y-axis (green), and z-axis (blue). 2) Despite additional visual cues, participants struggled with path accuracy due to depth perception limitations.
  }
    \label{fig:arc}
    \vspace{-1em}
\end{figure}

\textit{More Cues Does Not Mean Better Accuracy:} 
Despite the common use of path encoding in AR exercise research and the claimed importance of seeing path trajectories \cite{tang_physiohome_2015,yu_perspective_2020}, 
the path encoding showed similar---and, at times, worse---accuracy to other encodings for all exercises. 
This finding suggests that, after basic instruction, people may intuitively understand the expected motion trajectory from just the start and end positions. While more feedback may be necessary for teaching new movements, simple target-based designs 
were 
sufficient for single-limb motion guidance while minimizing potential negative effects from clutter. This finding supports Tang et al.'s recommendation to minimize excessive information \cite{tang_physiohome_2015}. Target-based visualizations also open opportunities to leverage the additional visual space to incorporate game-based elements \cite{powell_openbutterfly_2020, elor_project_2019} or progress feedback \cite{cavalcanti_usability_2019, bell_verification_2019,seyedebrahimi_brain_2019}. 

 Depth perception challenges potentially affected the accuracy of path encoding. Despite the presence of traditional VR depth cues such as 
 binocular disparity and prospective projection, participants sometimes struggled to align themselves with path visualizations in 3D space. P10 described this experience: 
 ``It looks like my hand is on the target based on the perspective, not sure if I am too forward or backwards.'' Figure \ref{fig:arc} illustrates this phenomena. From the participant's perspective, their hand appeared to follow the intended path (Figure \ref{fig:arc}.1). However, when we plot a representative participant's hand trajectory from above (Figure \ref{fig:arc}.2), 
 their hand consistently moved in front of the presented path. Multiple participants experienced this challenge regardless of encoding, which reflects people's tendency to underestimate depth in AR \cite{diaz_designing_2017,jones_effects_2008}. Unlike other encodings, paths intuitively encourage a sense of a ``correct'' motion execution, which may lead to greater false confidence when performing an exercise incorrectly.
Designs should consider 
potential depth challenges by creating  
larger or distorted target thresholds 
 or applying explicit distortions to account for perceptual bias in underestimating depth \cite{luboschik_spatial_2016}. Various visual cues could also mitigate this depth disparity, including drop shadows, 3D grids, and interactive progress feedback. 

Path encodings also resulted in slower movement, particularly for longer paths. This difference likely corresponds to participants 
focusing more on aligning with the path trajectory than on the target. Slower pace could be advantageous 
when controlled, deliberate motion is crucial for effectiveness or safety but 
may increase fatigue 
\cite{hansberger_dispelling_2017}. Pacing goals vary based on objectives. For example, 
resistance training typically employs slow, controlled movements to maximize muscular gains, while rehabilitative 
exercises may incorporate quick, dynamic movements such as arm swings and rapid punching motions to improve mobility and function. Future systems should consider individual needs 
to balance these trade-offs.

%% file: 7_Limitations_and_Future_Work.tex
\section{Limitations and Future Work}
\label{limitations}
Our study provides 
design insight for guiding visible 
motion with AR. We 
worked with physical therapists to select motions that reflect a range of movements employed in many real-world applications, work across planes, and could be addressed using 
both 1P and 3P. Future research should include a broader range of 
movements, including those with low visibility, 
multiple limbs, multiple planes of motion, and interactions with environmental elements,
\rev{to better understand the generalizability of our findings}.

Designing and evaluating visualizations 
to comfortably support wider range of movements may necessitate different display and capture hardware. 
Headset advances may also help capture and assess movement patterns. 
While 
the Vicon was beneficial for understanding precise motion patterns, 
it is impractical for real-world use, and affordable options like the Microsoft Kinect lack accuracy. Emerging egocentric motion capture technology shows promise for accurate, home-friendly body tracking \cite{zhang_reconstruction_2023}. Future hardware improvements could facilitate more practical movement tracking and guidance 
and enable insight into additional design variables, such as the impact of visualizations on long-term adherence. 
As real-world AR motion guidance becomes more feasible, future work can explore how these design recommendations are integrated in systems to improve long-term, task-specific outcomes.


While our study focused on using AR visualization to guide motion, 
future systems may 
explore how such visualizations mediate other forms of interaction, such as teaching new movements 
, providing direct 
instruction, \rev{or} 
remotely monitoring student progress and adjust targets and goals based on observed data. 

Motion guidance needs may vary across 
populations, such as the elderly, people with chronic conditions, or people with acute injury. 
While we focus on a general population to provide a broad foundation for movement preferences and patterns \rev{and protect participant safety, this focus 
does not necessarily capture the needs of specific patient groups, such as aging patients.} Future applications could extend our methods to 
specific populations \rev{and a larger sample size beyond those recommended by local standards in HCI research to increase statistical power \cite{caine2016local}. }

\rev{
Future work may 
investigate measures tailored to specific goals. For example, eye-tracking 
would capture focus and attention during exercises to see if 1P and 3P performance differences stem from attentional allocation or movement mechanics. Comfort, presence, and workload could help tailor designs for varying populations, such as those with cognitive or physical limitations. }

%% file: 8_Conclusion.tex
\section{Conclusion}
\rev{We offer expanded insight into guiding motions in AR} through an experiment examining the effects of encoding and perspective on accuracy and usability across three common isolation exercises varying in plane of motion and visibility. 
Our results demonstrate the efficacy of visualization techniques for guiding motion. From our findings, we suggest designers select optimal encodings and perspectives based on individual needs and motion visibility. This research contributes to the development of AR guidance systems that can improve motion guidance across domains, making health and training more accessible. 